\documentstyle[amssymb,twocolumn,prl,aps,graphicx]{revtex}
\begin{document}
\draft
\twocolumn[ \hsize\textwidth\columnwidth\hsize\csname@twocolumnfalse\endcsname
\title{
       Collective dynamics of liquid aluminum probed by \\
       Inelastic X-ray Scattering.
      }
\author{
        T.~Scopigno$^{1}$,
        U.~Balucani$^{2}$,
        G.~Ruocco$^{3}$,
        F.~Sette$^{4}$
        }
\address{
    $^{1}$Dipartimento di Fisica and INFM, Universit\'a di Trento, I-38100, Povo, Italy.\\
    $^{2}$Istituto di Elettronica Quantistica CNR, I-50127, Firenze, Italy. \\
    $^{3}$Dipartimento di Fisica and INFM, Universit\'a di L'Aquila, I-67100,L'Aquila, Italy.\\
    $^{4}$European Synchrotron Radiation Facility, B.P. 220 F-38043 Grenoble, Cedex France.
    }
\date{\today}
\maketitle

\begin{abstract}
An inelastic X-ray scattering experiment has been performed in liquid
aluminum with the purpose of studying the collective excitations at wavevectors 
below the first sharp diffraction peak.
The high instrumental resolution (up to 1.5 meV) allows an accurate
investigation of the dynamical processes in this liquid metal on the basis of
a generalized hydrodynamics framework. The outcoming results confirm the presence of 
a viscosity relaxation scenario ruled by a two timescale mechanism, as recently found 
in liquid lithium.
\end{abstract}


]

\section{INTRODUCTION\label{introduction}}

In the last two decades, the dynamical properties of liquid metals have been 
widely investigated aiming at the comprehension of the role of the mechanisms
underlying the atomic motions at the microscopic level. 
In particular, in the special case of alkali metals, it is known
that well-defined oscillatory modes persist well outside the strict
hydrodynamic region, down to wavelengths of a few inter-particle distances, making
these systems an ideal workbench to test the different theoretical
approaches developed so far for the microdynamics of the liquid state. 
By the experimental point of view, it is worth mentioning the pioneering 
inelastic neutron scattering (INS) experiment by Copley and Rowe \cite{rubidio2} 
in liquid rubidium, while more recently a lot of experimental efforts in 
performing more and more accurate experiments have been done: INS
investigations have been devoted to liquid cesium \cite{cesio}, sodium \cite{sodio}, 
lithium \cite{verkerk}, potassium \cite{potassio} and again rubidium \cite{rubidio}. 
Many numerical studies have been reported on the same systems, allowing 
to overcome experimental restriction such as the ($Q-E$) 
accessible region when dealing with the collective properties (with the density 
fluctuation spectra), and the simultaneous presence of coherent and incoherent contribution 
in the INS signal, the paramount experimental technique in this field up to a few years ago.
Moreover, numerical techniques allow to access a wider set of correlation function while 
the inelastic scattering experiment basically probe the density autocorrelation function.

On the theoretical side, tanks to the development of tools such as the 
memory function formalism, the relaxation concept, the kinetic theory, \cite{mori,lev} 
a framework has been established in order to describe the behavior of the afore mentioned 
correlation functions. In this respect, among the most important task, 
there is the idea that the decay of the density autocorrelation function 
occurs over different time scales. 
In particular a first mechanism is related to the coupling of the density and the 
temperature modes (thermal relaxation), while a second process involves the stress 
correlation function (viscosity relaxation). The latter mechanism proceeds through 
two different relaxation channels, active over different time scales: a first rapid 
decay, that has been ascribed to cage effects due to microscopic interaction of each 
atom and the cage of its neighboring atoms, followed by a slower process related 
to the long time rearrangements that drives the glass transition in those systems 
capable of supercooling. This structural relaxation process has been widely studied 
in the mode coupling formalism and recently remarkable efforts have been developed 
to set up self consistent approaches \cite{casasMC}.      
This theoretical framework has been tested on a number of numerical studies (see for 
example \cite{bal} and references therein). Indeed, the lineshape of the density correlation spectra 
extracted by INS (in those systems where it is allowed in the significant 
($Q-E$) region) did not allow to discriminate between different models including 
all the relevant relaxation processes at the microscopic level \cite{cesio}.

In the recent past, the development of new synchrotron radiation facilities
opened the possibility of using X-rays to measure the $S(Q,\omega)$ (which
is proportional to the scattered intensity) in the non-hydrodynamic region;
in this case the photon speed is obviously much larger than the excitations
velocity, and no kinematic restriction occurs. Moreover, in a monatomic
system, the Inelastic X-ray Scattering (IXS) cross section is purely 
coherent and so it is directly associated with the dynamic structure factor.

Recently, an accurate IXS study on liquid lithium allowed to 
detect experimentally the presence of the relaxation scenario 
predicted by theory on this prototypical alkali metal \cite{euro}. 
In particular, the data have been analyzed within a generalized hydrodynamic 
approach \cite{bal,boo} and, following a memory function formalism \cite{mori},
it has been possible to detect the presence of two relaxation processes
(additional to the thermal process resulting by the coupling of density and
temperature variables) affecting the dynamics of the systems in the
mesoscopic wavevector region \cite{prl}. This results, obtained with a
phenomenological ansatz for the memory function originally proposed in a
teorethical work by Levesque et al. \cite{lev}, has been then deeply
discussed pointing out the physical origin of such dynamical processes
\cite{scop}.

Although alkali metals are commonly referred to as the ``paradigm of
simple liquids'', other liquid metals always exhibit the characteristic 
structural and dynamical features that can be interpreted within the elementary 
approaches typically utilized when dealing with Lennard-Jones or alkali 
interatomic potentials \cite{earth,lead}. 
A remarkable example is provided by liquid aluminum. Several theoretical 
\cite{gask} and numerical \cite{ebb} works have been reported on this system. 
In fact, the basic common features of simple liquids, such as the presence of collective excitations
in the coherent dynamic structure factor below $Q_m$ and the positive
dispersion of the sound speed associated to them have been extensively
studied in this liquid. On the other hand, in this $Q$ region no experimental data have 
been reported to provide a real test and/or comparison of collective dynamics.
Indeed, in liquid Al, the high value of the isothermal sound speed $c_{0}\sim 4700$ m/s 
prevents for kinematical reasons a study of the collective dynamics by inelastic 
neutron scattering (INS): below $Q_m$ the available energy transfer is too 
limited to investigate the acoustic excitations properties. 
Accurate INS data have instead been reported at higher wavevectors where, 
however, the single particle response is mainly probed 
(see for example Ref. \cite{eder}).  
   
In this work we present the study of the coherent dynamic structure
factor, $S(Q,\omega )$, of liquid aluminum ($T=1000$ K) by Inelastic X-ray
Scattering, in a wavevector region, $0.05Q_{m}\leq Q\leq 0.5Q_{m}$, of
crucial importance as far as the dynamical features of the
collective motion are concerned.

\section{THE EXPERIMENT}
Our IXS experiment has been carried out at the high resolution beamline ID16
of the European Synchrotron Radiation Facility (Grenoble, F). The incident
beam is obtained by a back- scattering monochromator operating at the $(hhh)$
silicon reflections, with $h$$=$$9,11$. The scattered photons are collected 
by spherically bended Si crystal analyzers working at the same $(hhh)$ 
reflections. The total energy resolution function, measured from the elastic 
scattering by a Plexiglas sample, has a full width half maximum of 3 meV 
for $h$$=$$9$ and of 1.5 meV for $h$$=$$11$: this higher resolution 
configuration has been used at $Q=1$ and $2$ nm$^{-1},$ where the width
of the spectral features starts to be comparable to the resolution width at 
$h$=9. The wavevector transferred in the scattering process, $Q$$=$$%
2k_{i}sin(\theta _{s}/2)$ (with $k_{i}$ the wavevector of the
incident photon and $\theta _{s}$ the scattering angle) is
selected between 1 nm$^{-1} $ and 14 nm$^{-1}$ by rotating a 7 m
long analyzer arm in the horizontal scattering plane. The total
($FWHM$) $Q$ resolution has been set to 0.4 nm$^{-1}$. A five analyzers
bench was used to collect simultaneously five different $Q$
values, determined by a constant angular offset of 1.5$^{o}$
between neighboring analyzers. Energy scans have been performed by
varying the temperature of the monochromator with respect to that
of the analyzer crystals. Each scan took about 180 min, and each
spectrum at a given $Q$ was obtained from the average of 2 to 8
scans, depending on the values of $h$ and of $Q$. The data have
been normalized to the intensity of the incident beam. The liquid
aluminum sample (0.999\% purity) has been kept in an Al$_{2}
$O$_{3}$ container with optically polished single crystal sapphire
windows (0.25 mm thick). The sample length, which for an
IXS scattering experiment is optimal if coincident with the
absorption length, has been set to 1.0 mm in order to be nearly optimized
for both $h=9,11$ energies (21747 and 17794 eV respectively). 
The cell was then host into a molybdenum oven in thermal contact 
with a tantalum foil heated by the dissipation of about 100 W of 
power. All the environment has been kept in 10$^{-7}$ mbar dynamic vacuum.

The IXS spectra $I(Q,\omega)$, collected at fixed $Q$ as functions of the exchanged energy, 
are reported in Fig. 1. As $Q_m=25$ nm$^{-1}$, from the raw spectra we can 
observe the relevant region of dispersion of the acoustic mode. Initially, 
the frequency of the inelastic peaks increases linearly with $Q$. Despite 
an increase with $Q$ of their width, the modes are recognizable even at the 
highest wavevectors explored in the experiment. As expected, the ratio 
between the energy loss/gain sides of the spectra is ruled out by the 
detailed balance condition.
The quasi-elastic portion of the spectra, associated to some still unrelaxed 
damping mechanism, shows a linewidth that increases with $Q$, as the 
characteristic timescale of such process suddenly decreases at decreasing 
the dominant wavelenght of the observed density fluctuation.

At the lower Q-values a non-negligible contribution 
coming from the empty container scattering is clearly visible (arrows in Fig. 1). 
This feature basically stems from the inelastic scattering of the sapphire windows. 
As the longitudinal phonon velocity in Al$_2$O$_3$ is more than 10000 m/s, this 
``spurious'' contribution is beyond the energy region relevant for liquid Al and 
does not significantly affect the main spectral features of interest.
However, for an accurate quantitative assessment of the spectral shape 
at the lower $Q$ values, in the analysis we have excluded the undesired 
portions around the Al$_2$O$_3$ phonons by cutting off the two corresponding 
energy ranges (shadowed areas in Fig. 2).

\begin{figure}[]
\includegraphics[width=.6\textwidth]{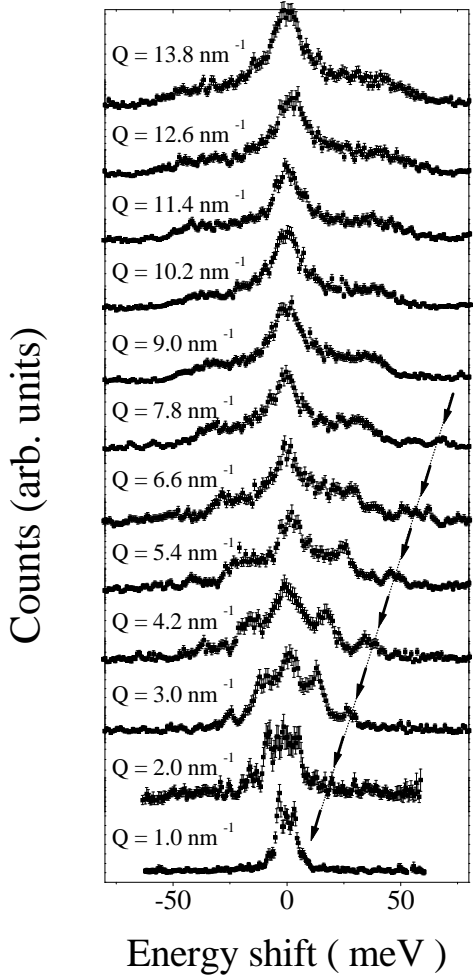}
\caption{
IXS spectra of liquid aluminum at the indicated wavevectors. 
The instrumental resolution is $\Delta E=1.5$ meV at $Q=$1 and 2 
nm$^{-1}$ and  $\Delta E=3.0$ meV elsewhere.
The total integration time is about 500 s for each point.}
\label{Fig1}
\end{figure}  

\section{DATA ANALYSIS}

The spectra $S(Q,\omega)$ 
have been described within the framework of the generalized Langevin equation 
for the density correlator $\phi (Q,t)=\left\langle \rho _{Q}(t)\rho _{-Q}(0)
\right\rangle / \langle \left|\rho_{Q}(t)\right|^{2}\rangle$:

\[\stackrel{..}{\phi }(Q,t)+\omega _{0}^{2}(Q)\phi (Q,t)+\mbox{$\int_0^t$}%
M(Q,t-t^{\prime })\stackrel{.}{\phi }(Q,t^{\prime })dt^{\prime }=0
\]
where
\begin{equation}
\omega _{0}^{2}(Q)=[k_{B}T/mS(Q)]Q^{2}=(c_{0}(Q)Q)^{2},
\label{sdq}
\end{equation}
In these equations, $m$ is the atomic mass, $S(Q)$ the static 
structure factor, $c_0(Q)$ the $Q$-dependent isothermal sound velocity and $M(Q,t)$ the
full memory function. Recalling that the Fourier transform of $\phi(Q,t)$
is $S(Q,\omega )/S(Q)$, and introducing the Fourier-Laplace transform of
the memory function as $M(Q,\omega )=M^{^{\prime \prime }}(Q,\omega )
+iM^{\prime }(Q,\omega )$, the previous expression becomes:
\begin{equation}
\frac{S(Q,\omega )}{S(Q)}=\frac{\pi ^{-1}\omega _{0}^{2}(Q)M^{\prime
}(Q,\omega )}{\left[ \omega ^{2}-\omega _{0}^{2}+\omega M^{\prime \prime
}(Q,\omega )\right] ^{2}+\left[ \omega M^{\prime }(Q,\omega )\right] ^{2}}
\label{sqw}
\end{equation}
All the details of the microscopic interactions 
are now embodied in the memory function $M(Q,t)$. For the latter, we have 
allowed a two-time relaxation mechanism of non-thermal contributions by 
assuming that

\begin{eqnarray}
\label{mem}
M(Q,t) &=&\left( \gamma -1\right) \omega _{0}^{2}(Q)e^{-D_{T}
Q^{2}t}  \\ &+&\Delta ^{2}(Q)\left[ A(Q)e^{-t/\tau _{\alpha
}(Q)}+(1-A(Q))e^{-t/\tau _{\mu }(Q)}\right]   \nonumber
\end{eqnarray}
where $\tau$'s, $\Delta ^{2}$ and $A$ are the timescales, the total 
viscous strength and the relative weight of the two processes respectively. 
In Eq. (\ref{mem}) the first term comes from the coupling between density and
thermal fluctuations ($\gamma=C_P/C_V $ is the specific heats ratio and $D_{T}=\kappa/nC_V$ 
the thermal diffusivity), while the other two contributions describe the relaxation of 
purely viscous decay channels. As is well known, for the latter the familiar 
viscoelastic model assumes a single exponential decay. However, as already 
found in liquid lithium \cite{prl,scop}, this simple ansatz is not accurate enough to reproduce 
the details of the IXS spectra in liquid Al.

Before any discussion of the fitting procedure, we recall that the actual 
scattered intensity is proportional to the convolution between the experimental 
resolution function and the true (quantum mechanical) dynamic structure factor 
$S_q(Q,\omega )$, affected by the detailed balance condition. To account for 
the latter, we have used the following approximation
\begin{equation}
S_{q}(Q,\omega )={\beta \hbar \omega }/(1-e^{-\beta \hbar \omega })
S(Q,\omega )  \label{kubo}
\end{equation}
which connects $S_q(Q,\omega )$ with its classical counterpart $S(Q,\omega )$.  

Summing up, we used Eq.(\ref{sqw},\ref{mem}) as a model function. We modified it 
according to Eq.(\ref{kubo}) and, finally, we folded it with the experimental 
resolution. The result has been utilized as fitting function $F(Q,\omega)$ for 
the scattered intensity $I(Q,\omega)$:
\begin{equation}
\label{fqw}
F(Q,\omega)=E(Q) \int{R(\omega-\omega')S_{q}(Q,\omega')d\omega'}
\end{equation}
where the constant $E(Q)$ depends on each analyzer efficiency and on the 
atomic form factor.   
The parameters $S(Q)$ and $\omega_0^2(Q)$ are related by the Eq. (\ref{sdq}), so 
that the most obvious procedure would be to put the data on an absolute 
scale estimating $S(Q)$ in a fitting-independent way and to fix $\omega_0^2(Q)$ 
accordingly. Consequently, {\it i)} the constant $E(Q)$ in Eq. (\ref{fqw}) 
would be $E(Q)=1$ {\it ii)} the only free fitting parameters would 
be the relaxation times and strengths of the viscous channels (for the thermal quantities 
$D_{T}(Q)$ and $\gamma(Q)$ it has  been utilized their $Q\rightarrow 0$ value). 
This method has been successfully applied in the lithium experiment, where 
the $S(Q)$ has been estimated from the raw spectra utilizing the first two sum 
rules corrected for finite resolution effects \cite{scop}.
In the present case instead, due to the presence of empty cell contributions 
in the tails of the scattered intensity, an estimate of the spectral moment would be not 
reliable. For this reason we have chosen to leave $E(Q)$ and $\omega_0^2(Q)$ 
as free parameters. The best-fitted value of the latter quantity has been 
then used to calculate $S(Q)$, putting the data on absolute scale.

\begin{figure}[h]
\centering
\includegraphics[width=.5\textwidth]{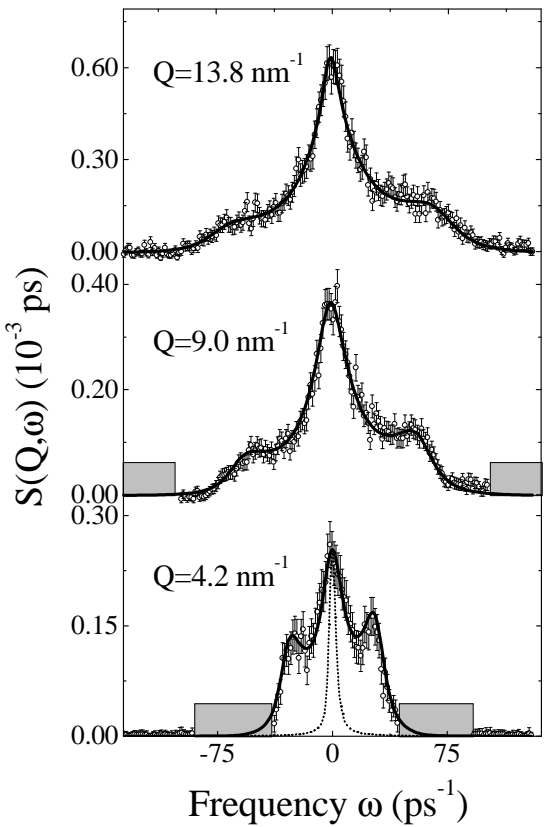}
\caption{
Selected examples of the fitting procedure. Open circle with the error bars
are the IXS data, the full lines are the best fit to the data. The shadow boxes 
indicate the energy windows that have been cut off 
in the fitting session. The resolution function is also reported (dotted line).
}
\label{Fig2}
\end{figure}

Some examples of the quality of the described fitting procedure are 
reported in Fig. 2, while the reliability of the normalization method 
is illustrated in Fig. 3, where the insert reports in a log scale 
the parameter $\omega_0^2(Q)$. The value of $S(Q)$ at small $Q$ is seen 
to be in fair agreement with the thermodynamic value  (arrow) \cite{ebb}.

In Fig. 4 the relative weight 
$A(Q)=\Delta_{\alpha}^2(Q)/(\Delta_{\alpha}^2(Q)+\Delta_{\mu}^2(Q))$ is shown. 
Compared to the liquid lithium, $A(Q)$ in liquid Al appears to 
have a faster decrease with the wavevector, indicating a somehow smaller role 
of slow features at finite wavevectors in this system. 

\begin{figure}[ftb]
\centering
\includegraphics[width=.5\textwidth]{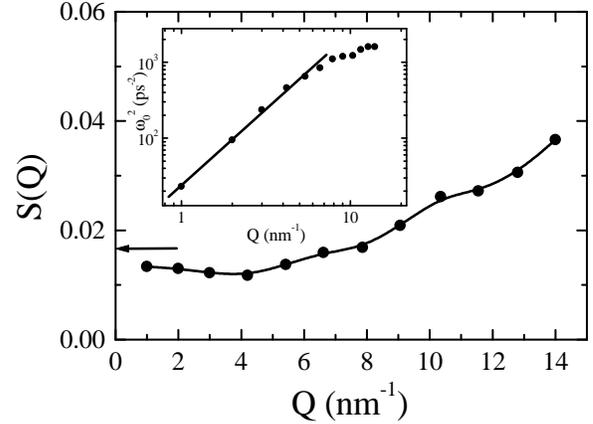}
\vspace{-7cm}
\caption{
Inset: $\omega_0^2(Q)$ as free fitting parameter. 
The full line is 
the expected power two behavior. Figure: Values of $S(Q)$ deduced by the previously 
determined $\omega_0^2(Q)$ as $S(Q)=KTQ^2/m\omega_0^2(Q)$. The arrow indicates the 
thermodynamic value of $S(0)$ by compressibility data.}
\label{Fig3}
\end{figure}

\begin{figure}[ftb]
\centering
\includegraphics[width=.44\textwidth]{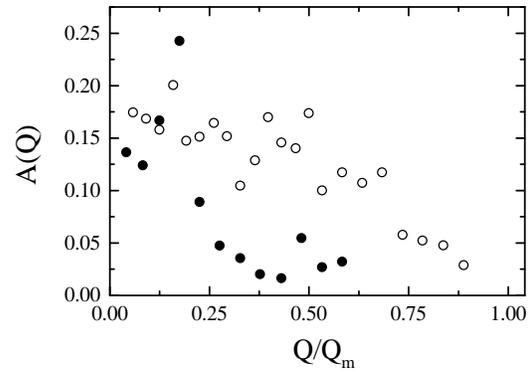}
\vspace{-5cm}
\caption{Ratio, $A(Q)$, between the slow and the total relaxation strength. Full dots
($\bullet $) lithium, open dots ($\circ $) aluminum, both slightly above the melting point. In order to compare the two systems, the exchanged wavevector has been normalized to the static structure factor maximum.
}
\label{Fig4}
\end{figure}

In Fig. 5 we report for both mechanisms the relaxation times and, in the 
inset, the quantity $\omega_l(Q)\tau(Q)$, where $\omega_l(Q)$ is the maximum of 
the longitudinal current correlation spectra.

\begin{figure}[ftb]
\centering
\includegraphics[width=.50\textwidth]{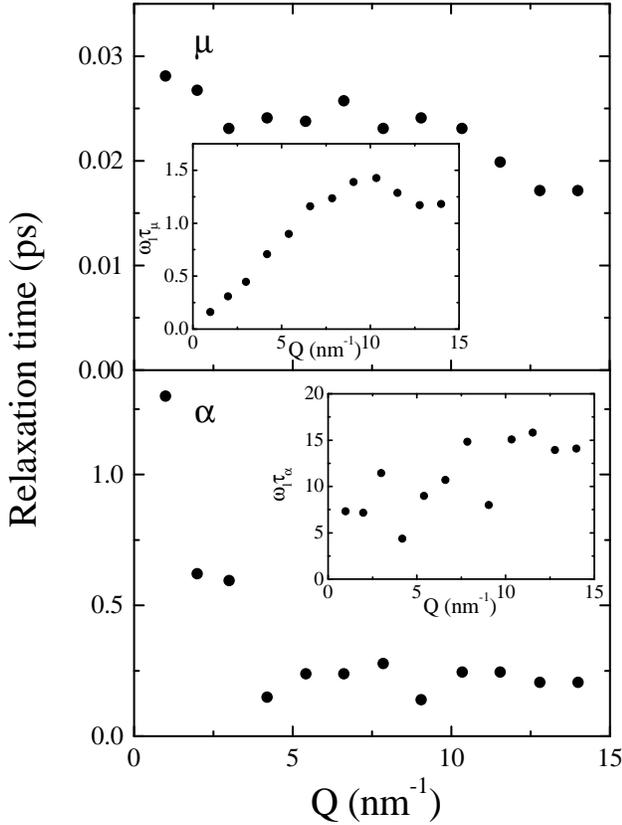}
\caption{
Relaxation times and $\omega_l(Q)\tau(Q)$ values determined by 
the fitting}
\label{Fig5}
\end{figure}
 
The characteristic time $\tau_\mu$ of the microscopic process 
moderately decreases with $Q$, while the much longer $\tau_\alpha$ is almost 
constant, except an abrupt initial decrease at very low $Q$ which may be 
possibly due to finite resolution effects. From some algebraic manipulations 
of Eqs. (\ref{sqw},\ref{mem}), it is possible to infer the relations between 
the parameters ruling each mechanism and the main spectral features, i.e. the 
peak position and the width of the acoustic mode, as well as the intensity and 
the width of the central quasi-elastic contribution.
From the inset of Fig. 5 it can be seen that for the slow process the condition 
$\omega_l(Q)\tau_\alpha(Q)>1$ always holds, so that this mechanism contributes mostly 
to the elastic linewidth rather than to the acoustic mode broadening.
On the other hand, the fast microscopic process is such that $\omega_l(Q)\tau_\mu(Q)\leq1$ 
at low $Q$, so that in this region it contributes to the acoustic damping. 
However, at wavevectors $Q>5$ nm$^{-1}$, one reaches the crossover 
$\omega_l(Q)\tau_\mu(Q)\approx 1$, so that the fast mechanism drives an effective 
increase of the sound speed toward a limiting, solid-like, instantaneous response. 
As the weight of the fast process exceeds the one of the slow one (namely 
$A(Q)<<1$ in Eq. (\ref{mem}), cf. Fig. 4), the increase of the sound speed 
basically stems only from the fast microscopic mechanism, which above the 
crossover threshold is also responsible for the broader tails of the quasi-elastic 
portion of the spectra ($\tau_{\alpha}>>\tau_{\mu}$).

In Fig. 6 we report the {\it deconvoluted} spectra, i.e. the $S(Q,\omega)$ as obtained 
from the fitting procedure, together with the relative current correlation 
spectra $J(Q,\omega)=\omega^{2}/Q^{2}S(Q,\omega)$. The latter quantity plays 
an important role in the description of the dynamical features of a systems because 
its maxima are related to $c_{l}=\omega_{l}/Q$, namely the apparent sound 
speed of the system.

\begin{figure}[ftb]
\includegraphics[width=.50\textwidth]{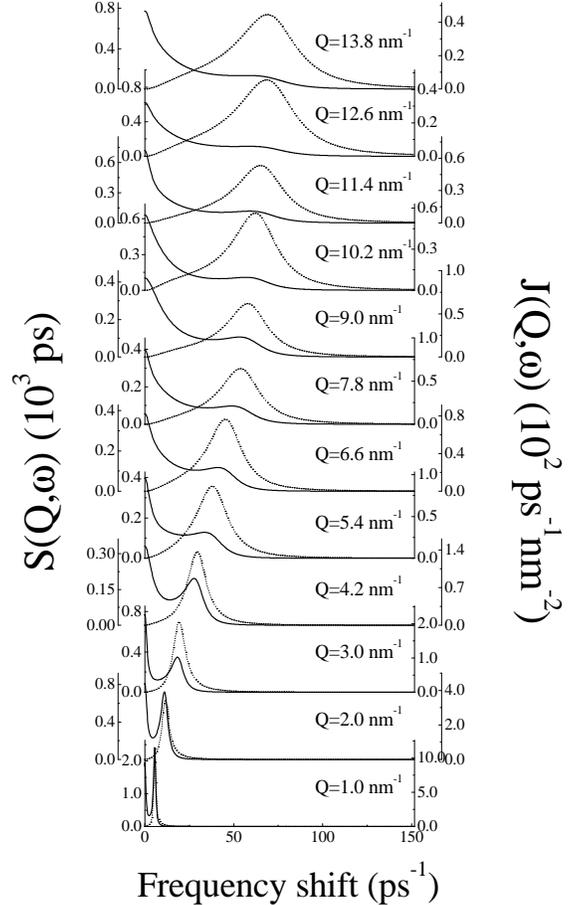}
\caption{
Deconvoluted spectra of the density correlation function $S(Q,\omega)$ (left axis) 
and of the current correlation function $J(Q,\omega)$ (right axis) as obtained 
from the fit.
}
\label{Fig6}
\end{figure}  

Finally, in Fig. 7, we compare the $Q$-dependent sound speed in liquid Al 
(calculated as the maximum of the fitting-deconvoluted longitudinal current 
correlation function $J(Q,\omega)=\omega^2/Q^2S(Q,\omega)$) with the one 
previously determined in liquid lithium. Suitable reduced units (velocities 
normalized to their respective isothermal values, wavevectors measured in 
units $Q_m$) have been adopted. 

\begin{figure}[ftb]
\centering
\includegraphics[width=.49\textwidth]{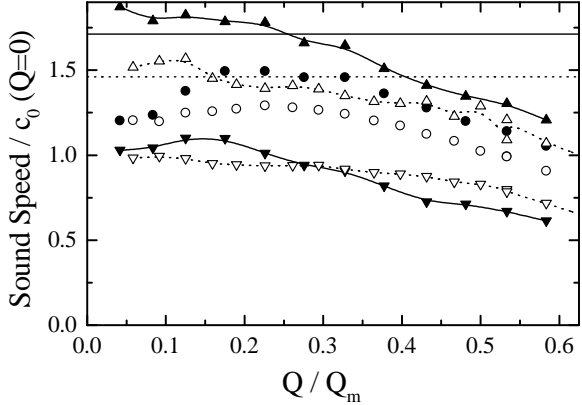}
\vspace{-7cm}
\caption{
Effective sound speed (dots) of aluminum (full symbols) vs lithium 
(open symbols). Data have been scaled by $Q_m$ -the first sharp diffraction peak and 
$c_0(0)$ -the isothermal speed of sound in the hydrodynamic limit ($c_{0}^{Li}=4450 m/s; c_{0}^{Al}=4700 m/s$).
$c_0(Q)$ is also reported ($-\triangledown-$) together with the $c_{\infty }(Q)$ 
values determined by the fitting ($-\vartriangle-$) and by the pair distribution 
function and interatomic potential (lines).
The lines connecting the symbols are guidelines for the eyes only.}
\label{Fig7}
\end{figure}

Both systems are seen to exhibit a clear 
positive dispersion effect, 
i.e. an increase of the sound speed at increasing the exchanged wavevector. 
Such dispersion, in ordinary liquids, proceeds between the adiabatic 
($Q\rightarrow 0$) limit, $c_s=\sqrt{\gamma}c_0$, and the unrelaxed, instantaneous 
value $c_\infty(Q)=\sqrt{\gamma\omega_{0}^{2}(Q)+\Delta^{2}(Q)}/Q$, where $\Delta^2$ 
is the total strength of the non-thermal processes. In metallic liquids, instead, 
due to the high thermal conductivity, the condition $\omega_l(Q)<D_{T}Q^{2}$ 
holds for $Q\gtrapprox 0.1 nm^{-1}$, i.e., in the IXS $Q$-window ($Q>1nm^{-1}$), 
no region exists where the three relaxation processes thermal, $\alpha$, $\mu$, 
are simultaneously unrelaxed. 
As a consequence, the lower and upper value of the sound speed are expected 
to be the isothermal $c_{0}(Q)$ (defined in Eq.(\ref{sdq})) and the partially 
unrelaxed $c'_\infty=\sqrt{\omega_{0}^{2}(Q)+\Delta_\alpha^{2}(Q)+\Delta_\mu^{2}(Q)}/Q$ \cite{nota}.

To test the reliability of our model, we finally compared $c_\infty^{FIT}(Q)$ 
with the theoretical value $c_\infty^{TH}(0)$ deduced from the structure 
and interatomic potential data \cite{gask,scopmd}.
In both systems the fitted values of $c_\infty^{FIT}(Q\rightarrow 0)$ slightly exceed the data for $c_\infty^{TH}(0)$ reported in literature. 
A possible explanation of this inconsistency may be ascribed to 
the arbitrary choice of the memory function shape, that in principle 
can have more complicated features than the double exponential ansatz 
of Eq. (\ref{mem}). 
The major drawback of such an assumption is, indeed, the cusp at $t=0$. 
It is reasonable to think that an exponential
decay forced to represent a more complicated time dependence can give an
accurate estimate as far as the $\tau $ is concerned, while at short times
the lack of a zero second derivative inescapably leads to an overestimate of
$M_L(Q,t=0)$, the positive dispersion amplitude. A similar effect,
i.e. the overestimation of the $c_\infty (Q)$ deduced by the fit has been also 
observed in liquid lithium, where the same memory function has been 
adopted \cite{scop}.

\section{CONCLUSION}

In this work inelastic X-ray scattering technique has been utilized to study in
detail the main features of the microscopic dynamics in liquid aluminum in the
mesoscopic momentum region ($Q\approx1-15$ nm$^{-1}$), i.~e. the  region were 
the collective properties are dominant. Aluminum is a system where the 
investigation of the dynamics using the well-established  inelastic neutron 
scattering technique is not possible. The present IXS experiments allowed
for the determination of the pattern of relaxation processes entering in the
density-density memory function, and therefore affecting the spectral shape of 
the dynamic structure factor. Specifically, the high quality of the data 
allows to identify univocally the presence of three distinct relaxation 
processes. The first one -the usual thermal relaxation- is associated with the 
coupling between density and energy fluctuations. As in alkali metals, although
its relevance is rather small, this process cannot be neglected; more interesting,
it plays a role which is different with respect to the ordinary non-conducting 
liquids. Much more important is the unambiguous evidence of {\it two well 
separated timescales} in the decay of that part of the memory function which 
id associated to the generalized viscosity. In particular, the present experiment
proves that the traditional description of the generalized viscosity by a single 
time-scale (viscoelastic model) cannot account for the detailed shape of 
$S(Q,\omega )$, and that it only provides a qualitative description of the 
microscopic dynamics  in liquid metals.

The presence of a two time-scales decay of the memory function poses
a question of crucial importance: namely, their physical origin and, most 
important, of the fast one.

The slow process, making use of the terminology used to describe glass
forming systems, is related to the $\alpha $-relaxation which, in systems 
capable to sustain strong supercooling, is responsible for the liquid-glass 
transition. From a more canonical point of view, this process can be framed 
within kinetic theory in terms of ''correlated collisions'': in mode coupling 
approaches, the onset of these correlation effects is traced back to the coupling 
to slowly relaxing dynamical variables, and specifically in the liquid region, 
to long lasting density fluctuations. The quantitative description of this
''slow'' timescale requires a full evaluation of the mode-coupling
contribution at different wavevectors which is beyond th aim of the present
work.

As far as the fast process is concerned, the situation is more confused and 
its interpretation is still matter of debate. This is particularly annoying 
because -as shown in this paper for liquid aluminum and as demonstrated in
the case of liquid lithium- the fast process is indeed more relevant than the
slow one: it largely controls both the sound velocity dispersion and 
attenuation in the mesoscopic $Q$ region. Historically, within 
a generalized  kinetic theory approach, the fast initial decay of the density 
fluctuation  memory function, is traced back to {\it collisional events}, which 
are fast, short range and, more important, uncorrelated among each others. 
Within this scheme, the short timescale $\tau_\mu$ turns out to be associated 
with the duration of a rapid structural rearrangement occurring over a spatial 
range $\cong 2\pi /Q_m$. Although this description of the fast process in term 
of uncorrelated interparticle collision is a possible way to qualitative account 
for the dynamical features at short times in liquids and dense systems, it 
seems to be unable to account for similar results obtained in glasses 
\cite{scopmd,glasslett,gotze}, where the {\it collective} aspect of the 
dynamics can not be neglected.

A possible, and different, approach to describe the initial fast decay of the 
memory function in dense fluids and liquids relies on the normal
modes (''instantaneous'' in normal liquids) analysis of the atomic dynamics
\cite{keyes}, an approach that works correctly in the limiting case of
"harmonic glasses". In this case one uses a framework (dynamical matrix, etc.)
formally similar to the one customarily adopted for harmonic crystals;
however, owing to the lack of translational symmetry of the system, it turns
out \cite{glasslett} that the eigenstates cannot anymore be represented by
plane waves (PW) even at relatively small wavevectors \cite{mazza}. 
A scattering experiment, where $Q$ is fixed, is equivalent -via the
fluctuation-dissipation relation- to a response experiment, where one study 
the time evolution of the system after a {\it sinusoidal} perturbation (with
period $2\pi/Q$) is applied to the system itself. In the case of measuring
the $S(Q,\omega)$, the applied perturbation is a density fluctuation.
As the PW are not eigenstate of the disordered systems, the initial
perturbation can be projected along different eigenmodes, each one evolving
in time with its own frequency. As a consequence of this frequency spread the 
energy initially stored in the PW {\it relax} towards other PW's with different
$Q$ values. This process is exactly what one expect in presence of a 
relaxation process, and its characteristic time is determined by the
projection of the system eigemodes on the PW basis (Fourier transform
of the eigenvectors). We deal, therefore, with a mechanism whose ultimate 
origin is the topological disorder, and not a truly dynamical event. Ordinary 
liquids (such as the one considered here) are certainly ''disordered systems'' and
at the high frequency considered here ($\omega \tau _\alpha >>1$), they can
be considered as ''frozen''; therefore the description used for glasses can
be applied as well. Obviously, the "harmonic approximation" and the residual 
effect of the finite value of $\omega\tau$ prevent the "quantitative"
application of the previously described formalism to the case of normal liquids.
However we expect that -on a qualitative ground- the fast relaxation process
in liquids can be described in term of "high frequency vibrations" (that are a 
strongly correlated atomic motion), and this point of view appears to be in 
contrast with the (uncorrelated) binary collision based description.

At the present stage the experimental data cannot support either interpretation.
However, the possibility that the phenomenology reported here for a simple liquid 
can be understood within the same mental framework developed for more complex 
liquids and glasses, is certainly appealing.


\begin{references}
\bibitem{nota} By a numerical point of view, the difference between 
$c_\infty(Q)$ and $c'_\infty(Q)$ is of the order of $\sqrt\gamma$ ($3\%$), 
namely, within the statistical errors of the fitting procedure. 
It is worth to point out how the thermal process, contrarily to 
what happens in ordinary liquids, gives also a non-viscous damping 
$\delta \omega =\left( \gamma -1\right) c_{0}^{2}(Q)/D_T$, responsible 
for an additional Brillouin linewidth particularly important at the lower 
$Q$'s of our experiment.
\end{references}
\end{document}